\documentclass[astrosymb,twocolumn]{aastex701}

\newcommand{\kms}{km\,s$^{-1}$}
\newcommand{\mjybeam}{mJy\,beam$^{-1}$}
\newcommand{\mujybeam}{$\mu$Jy\,beam$^{-1}$}
\newcommand{\msun}{${\rm M}_\sun$}

\newcommand{\methanol}{CH$_3$OH}
\newcommand{\metcyn}{CH$_3$CN}
\newcommand{\metcold}{$E\,v_t=0\,J_{K_a,K_c}=8_{-0,8}-7_{-1,6}$}

\defcitealias{olguin_digging_2021}{DIHCA I}
\defcitealias{olguin_digging_2022}{DIHCA II}
\defcitealias{taniguchi_digging_2023}{DIHCA III}
\defcitealias{ishihara_digging_2024}{DIHCA IV}
\defcitealias{sakai_digging_2025}{DIHCA V}
\defcitealias{luo_digging_2026}{DIHCA VI}
\defcitealias{olguin_digging_2026}{DIHCA VII}

\begin{document}

\title{Digging into the Interior of Hot Cores with ALMA (DIHCA). VIII. Feeding a high-mass binary system in the making}

\author[0000-0002-8250-6827]{Fernando A. Olguin} 
\email[show]{f.olguin@yukawa.kyoto-u.ac.jp}
\affiliation{Center for Gravitational Physics, Yukawa Institute for Theoretical Physics, Kyoto University, Kitashirakawa Oiwakecho, Sakyo-ku, Kyoto 606-8502, Japan}
\affiliation{National Astronomical Observatory of Japan, National Institutes of Natural Sciences, 2-21-1 Osawa, Mitaka, Tokyo 181-8588, Japan}

\author[0000-0002-0197-8751]{Yoko Oya}
\email{yoko.oya@yukawa.kyoto-u.ac.jp}
\affiliation{Center for Gravitational Physics, Yukawa Institute for Theoretical Physics, Kyoto University, Kitashirakawa Oiwakecho, Sakyo-ku, Kyoto 606-8502, Japan}

\author[0000-0002-7125-7685]{Patricio Sanhueza} 
\email{psanhueza@astron.s.u-tokyo.ac.jp}
\affiliation{Department of Astronomy, School of Science, The University of Tokyo, 7-3-1 Hongo, Bunkyo, Tokyo 113-0033, Japan}

\author[0000-0002-6752-6061]{Kaho Morii}
\email{kaho.morii@cfa.harvard.edu}
\affiliation{Center for Astrophysics $|$ Harvard \& Smithsonian, 60 Garden Street, Cambridge, MA 02138, USA}

\author[0000-0003-1275-5251]{Shanghuo Li}
\email{shli@nju.edu.cn}
\affiliation{School of Astronomy and Space Science, Nanjing University, Nanjing, China}
\affiliation{Key Laboratory of Modern Astronomy and Astrophysics, Nanjing University, Ministry of Education, Nanjing, China}

\author[orcid=0000-0003-4402-6475]{Kotomi Taniguchi}
\email{kotomi.taniguchi@nao.ac.jp}
\affiliation{National Astronomical Observatory of Japan, National Institutes of Natural Sciences, 2-21-1 Osawa, Mitaka, Tokyo 181-8588, Japan}
\affiliation{Astronomical Science Program, The Graduate University for Advanced Studies, SOKENDAI, 2-21-1 Osawa, Mitaka, Tokyo 181-8588, Japan}

\author[0000-0002-9774-1846]{Huei-Ru Vivien Chen}
\email{hchen@phys.nthu.edu.tw}
\affil{Institute of Astronomy and Department of Physics, National Tsing Hua University, Hsinchu 30013, Taiwan}

\author[0000-0002-8691-4588]{Yu Cheng}
\email{ycheng.astro@gmail.com}
\affiliation{National Astronomical Observatory of Japan, National Institutes of Natural Sciences, 2-21-1 Osawa, Mitaka, Tokyo 181-8588, Japan}

\author[0000-0003-4506-3171]{Qiuyi Luo} 
\email{luoqiuyi233@gmail.com}
\affiliation{Department of Astronomy, School of Science, The University of Tokyo, 7-3-1 Hongo, Bunkyo, Tokyo 113-0033, Japan}
\affiliation{National Astronomical Observatory of Japan, National Institutes of Natural Sciences, 2-21-1 Osawa, Mitaka, Tokyo 181-8588, Japan}

\author[0000-0003-2384-6589]{Qizhou Zhang}
\email{qzhang@cfa.harvard.edu}
\affiliation{Center for Astrophysics $|$ Harvard \& Smithsonian, 60 Garden Street, Cambridge, MA 02138, USA}

\author[0000-0003-4521-7492]{Takeshi Sakai}
\email{takeshi.sakai@uec.ac.jp}
\affiliation{Graduate School of Informatics and Engineering, The University of Electro-Communications, Chofu, Tokyo 182-8585, Japan.}

\author[0009-0007-4384-6507]{Kongcheng Liu}
\email{kong715@g.ecc.u-tokyo.ac.jp}
\affiliation{Department of Astronomy, School of Science, The University of Tokyo, 7-3-1 Hongo, Bunkyo, Tokyo 113-0033, Japan}

\author[0009-0004-8232-4590]{Yutaka Nakazawa}
\email{nakazawa-yutaka114@g.ecc.u-tokyo.ac.jp}
\affiliation{Department of Astronomy, School of Science, The University of Tokyo, 7-3-1 Hongo, Bunkyo, Tokyo 113-0033, Japan}

\begin{abstract}
We analyze the gas kinematics in the proto-binary NGC 6334I(N) ALMAe9-ALMAe14 system with the Atacama Large Millimeter/submillimeter Array (ALMA) at 1.33\,mm with an angular resolution of 66\,mas (${\sim}$90\,au).
With a separation of 740\,au between the system members, these observations allow a detailed study of the gas motion around this likely binary system hosting at least one growing high-mass star.
Some \methanol\ lines tracing cold gas ($E_u/k_B{\sim}100$\,K) reveal the presence of ordered circumbinary gas motion around the binary system within a ${\sim}1000$\,au radius, and rotation around the individual sources with rotation axes pointing roughly in the same direction.
We estimate a central source mass, uncorrected for inclination, of 2\,\msun\ for the secondary, while the primary has an estimated mass of 5\,\msun\ as reported in previous works.
The \methanol\ transitions with higher $E_u/k_B$ ($\gtrsim400$\,K) values reveal circumstellar gas rotation in ALMAe9 consistent with the cold gas tracers.
The vibrationally excited SO emission also traces rotation around ALMAe9 and potentially an accretion shock from  a flow feeding the circumstellar disk of ALMAe14. 
We identify two potential inflows feeding the circumstellar disks from the circumbinary gas at rates of ${\sim}10^{-6}-10^{-4}$\,\msun\,yr$^{-1}$ from \methanol\ emission.
An additional third inflow seems to be feeding the primary from gas at larger scales.
\end{abstract}


\section{Introduction}\label{sect:intro}

Most stars are born in multiple systems, with high-mass stars in binary systems an important source of some of the most powerful phenomena detected across cosmic scales (e.g., X-ray binaries).
The formation of such systems at early stages can be the result of disk \citep[e.g.,][]{kratter_role_2010,mignon-risse_collapse_2021} or core \citep[e.g.,][]{krumholz_radiation-hydrodynamic_2007} fragmentation, with the former likely resulting in closer binaries than the latter.
Binaries formed through disk fragmentation can develop a circumbinary disk that feeds their individual circumstellar disks.
Simulations suggest that such disks would feed the secondary at a higher accretion rate depending on the core angular momentum \citep[][]{matsumoto_structure_2019}.
However, investigating how proto-binaries feed requires high-angular resolution observations in order to resolve the kinematics of the circumbinary and circumstellar disks.

At radio frequencies, this can be achieved with interferometers like ALMA.
For instance, ALMA has revealed inflows in the low-mass binary system [BHB2007] 11 feeding the individual circumstellar disks \citep[][]{alves_gas_2019}.
A larger spiral arm (streamer) has been found feeding the low-mass binary SVS13A \citep[][]{hsieh_prodige_2023}, while spiral arms within a circumbinary disk have been identified in L1551 \citep{takakuwa_circumbinary_2020}.
A circumbinary/multiple disk has also been detected around the hot corino IRAS 16293--2422 \citep[][]{oya_substructures_2020}.
On the higher mass end, circumbinary disks have been identified through line transitions tracing cold gas with molecules like CH$_3$CN and CH$_3$OH in the high-mass sources IRAS 16547--4247 \citep[G343.12--0.06, 2.9 kpc; e.g.,][]{tanaka_salt_2020,zapata_asymmetric_2019} and G335.579--0.272 ALMAe1 \citep[3.25 kpc;][]{olguin_digging_2022} within 1000\,au scales.
Other systems forming binaries include G11.92--0.61 MM1 \citep[3.37\,kpc; e.g.,][]{ilee_g1192-061_2018} and MM2 \citep[][]{zhang_filamentary_2024,sanhueza_magnetic_2025}, IRAS 07299--1651 \citep[1.68\,kpc;][]{zhang_dynamics_2019,wang_eccentric_2026}, G345.4938+01.4677 \citep[2.38\,kpc;][]{guzman_photoionized_2020} and multiple binary systems within G333.23-0.06 \citep[5.2\,kpc;][]{li_observations_2024}.
However, they could not identify gas flows feeding the circumstellar disks.

Of the few nearby massive sources, NGC 6334I(N) (G351.444+00.659) is a young cluster-forming region located close enough \citep[1.35 kpc;][]{reid_trigonometric_2014} to resolve and study multiple systems with ALMA without  using the more time consuming extended baseline configurations.
Within NGC 6334I(N), \citet{li_detection_2025} found a septuple system in the making.
Although their observations revealed molecular line emission toward the individual members of the system, they could not study the motions close to each source due to lack of enough angular resolution to separate the individual components.
As part of the Digging into the Interior of Hot Cores with ALMA (DIHCA) project, \citet[][]{olguin_digging_2026} detected line emission toward 5 sources within the field.
Of these, two (ALMAe9 and ALMAe14) are located within 1000 au, thus making them good candidates for a binary system.
However, due to the lack of sensitivity (in part due to their selection of a homogeneous robust parameter across the sample), gas rotation was only studied toward ALMAe9 resulting in a estimated central source mass of $M_c \sin^2 i=5$\,\msun, with $i$ the inclination of the putative disk with respect to the plane of the sky.

Here we present new observations at higher angular resolution with ALMA and a re-analysis of the kinematics in the NGC 6334I(N) ALMAe9-ALMAe14 system that reveal  circumbinary gas with potential inflows feeding the circumstellar disks.
The observations are presented in Section~\ref{sect:obs}.
The results are presented and discussed in Section~\ref{sect:results}, while the conclusions are presented in Section~\ref{sect:conclusions}.

\section{Observations}\label{sect:obs}

\begin{deluxetable}{lccc}
\tablecaption{Summary of observations.\label{tab:obs}}
\tablewidth{\textwidth}
\tablehead{
\colhead{Configuration} & \colhead{Date} & \colhead{Antenna} & \colhead{Baseline} \\
\colhead{Number}        & \colhead{}     & \colhead{Number}  & \colhead{range} \\
\colhead{}              & \colhead{}     & \colhead{}        & \colhead{(m)}
}
\startdata
C-5 & 2017 Jul 06 & 44 & 17-2647 \\
C-5 & 2017 Dec 31 & 44 & 15-2516 \\
C-8 & 2019 Jul 30 & 41 & 92-8548 \\
C-9 & 2023 Jul 20 & 45 & 230-15238 \\
\enddata
\end{deluxetable}

The NGC 6334I(N) region was observed with the ALMA 12\,m array at 1.33\,mm (band 6) as part of DIHCA and follow up observations.
Table~\ref{tab:obs} summarizes the observations per 12\,m array configuration, hereafter we will refer to each configuration by its configuration number.
Details of the DIHCA observations (C-5 and C-8) are described in \citet{ishihara_digging_2024,luo_digging_2026,olguin_digging_2026}, while the follow up observations (C-9) are explained in \citet{olguin_massive_2025}.
The spectral coverage and phase centers of the observations are slightly shifted, because the C-9 observations were designed to cover hotter lines and the positions were taken from the ATLASGAL survey \citep{urquhart_atlasgal_2018}.
The C-9 observations were calibrated using the corresponding Common Astronomy Software Applications \citep[CASA;][]{casa_team_casa_2022} pipeline version, and the continuum visibilities were produced using GoContinuum \citep[version v2.0]{olguin_gocontinuum_2020}.
Self-calibration was only applied to the continuum visibilities of the C-5 and C-8 observations.

As pointed out in \citet{olguin_digging_2026}, the combined continuum subtracted visibilities for NGC 6334I(N) produce negative bowls around the main sources once imaged.
Therefore, we adopt a similar strategy to produce the data cubes by imaging the non-continuum subtracted visibilities, and subtract the continuum using \textsc{statcont} \citep{sanchezmonge_statcont_2018}.
We use the CASA task \textsc{tclean} to image the combined visibilities with the mosaic gridder and Hogbom deconvolver with Briggs weighting and a robust parameter value of 1.
This robust value allow for a resolution similar to that of the C-8 alone but provides a higher sensitivity and recover more extended emission, which are lacking in the data cubes presented in \citet[][robust 0.5]{olguin_digging_2026}.
To produce the clean mask, we use the built-in auto-masking of the \textsc{tclean} task.
The resulting cubes have synthesized beams ranging between 55 and 69\,mas (75--93\,au), a spectral resolution of ${\sim}976.6$\,kHz${\sim}1.3$\,\kms, and noise levels of ${\sim}1.2$\,\mjybeam\ (${\sim}6$\,K) per channel.

\begin{figure}[ht]
\plotone{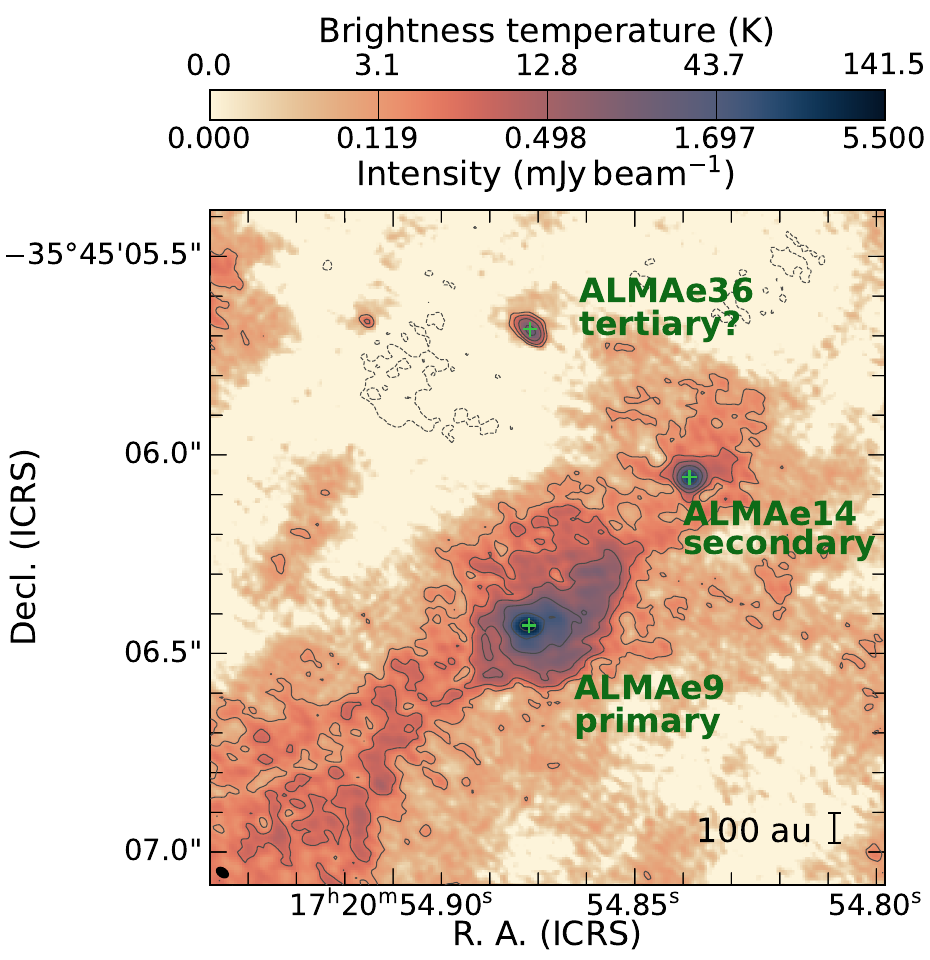}
\caption{NGC 6334I(N) ALMAe9 and ALMAe14 continuum emission at 1.33\,mm.
The contour levels are --6, --3, 3, 6, 12 ... $\times\sigma_{\rm rms}$, with $\sigma_{\rm rms}=41$\,\mujybeam\ the rms noise level.
The green crosses indicate the positions of the continuum sources.
The synthesized beam is shown in the lower left corner.
\label{fig:continuum}}
\end{figure}

Finally, we use the combined continuum visibilities to produce the continuum image.
We use the same parameters of the task \textsc{tclean} as those used to produce the data cubes, but with a robust parameter of 0.5.
The resulting continuum image presented in Figure~\ref{fig:continuum} have a synthesized beam of $0\farcs037\times0\farcs025$ and a noise level of 41\,\mujybeam.
A view of the complete NGC 6334I(N) field is shown in Appendix~\ref{app:figs}.

\section{Results and Discussion}\label{sect:results}

\subsection{Source morphology}\label{sect:morphology}

The continuum emission in Figure~\ref{fig:continuum} shows three bright sources within the region ALMAe9, ALMAe14 and ALMAe36.
The brightest source, ALMAe9, has a central peak surrounded by extended emission in part resembling a spiral arm to the north-west and clumpy substructures.
Low brightness emission between the 3 and 6$\sigma$ levels connects ALMAe9 and ALMAe14, while even fainter emission connects ALMAe14 and ALMAe36.
To the south-east, extended emission seems to connect ALMAe9 with the larger cloud.
The sources ALMAe9 and ALMAe14 are separated by 0\farcs55 (${\sim}740$\,au), thus likely forming a binary system (Luo et al. in prep.).
We fit a 2-D Gaussian to the primary beam corrected continuum data using CARTA \citep[version 5.1;][]{wang_cartacube_2026,angus_comrie_carta_2026}.
For ALMAe9 and ALMAe14, we use two components fitted simultaneously to account for background extended emission, while for the more isolated source ALMAe36 a single component is used.
The fitting results are listed in Table~\ref{tab:src_props}.
The only evidence supporting that ALMAe36 belongs to the ALMAe9-ALMAe14 system is the aforementioned faint ($<3\sigma$) continuum emission.
Therefore in what follows we will focus the analysis mainly in ALMAe9 and ALMAe14 and treat the pair as a (potential) binary system.

\begin{deluxetable*}{lcCCCCCCCCCCC}
\tablecaption{Source properties from continuum and line fitting.\label{tab:src_props}}
\tablehead{
\colhead{Source} & \colhead{Comp.} & \colhead{R. A.} & \colhead{Decl.} & \colhead{FWHM} & \colhead{P. A.} & \colhead{$R$} & \colhead{$I_\nu$} & \colhead{$F_\nu$} & \colhead{$T_B$} & \colhead{$T$} & \colhead{$M_d$} & \colhead{$\tau$} \\
\colhead{} & \colhead{}    & \colhead{([h:m:s])} & \colhead{([d:m:s])} & \colhead{(mas)} & \colhead{(degree)} & \colhead{(au)} & \colhead{(mJy beam$^{-1}$)} & \colhead{(mJy)} & \colhead{(K)}  & \colhead{(K)}  & \colhead{(\msun)} & \colhead{} \\
\colhead{} & \colhead{(1)} & \colhead{(2)} & \colhead{(3)} & \colhead{(4)} & \colhead{(5)} & \colhead{(6)} & \colhead{(7)} & \colhead{(8)} & \colhead{(9)} & \colhead{(10)} & \colhead{(11)} & \colhead{(12)}\\
}
\startdata
ALMAe9     & 1 & 17:20:54.87192 & -35:45:06.4296 & 83\times62   & -80 &  48 & 3.4 & 18.9 & 140     & 260 & 0.2 (0.3)  & 0.4 \\
           & 2 & 17:20:54.86673 & -35:45:06.4021 & 397\times253 & -48 & 210 & 0.9 & 93.0 & \nodata & 200 & 1.2 (1.6)  & 0.1 \\
ALMAe14    & 1 & 17:20:54.83867 & -35:45:06.0566 & 44\times41   & -92 &  29 & 3.0 &  5.8 & 84      & 160 & 0.1 (0.2)  & 0.7 \\
           & 2 & 17:20:54.83588 & -35:45:06.0347 & 239\times186 & -72 & 140 & 0.3 & 15.3 & \nodata & 140 & 0.3 (0.5)  & 0.06 \\
ALMAe36    & 1 & 17:20:54.87167 & -35:45:05.6838 & 59\times38   &  49 &  32 & 1.3 &  3.1 & 32      & 130 & 0.07 (0.3) & 0.3 \\
\enddata
\tablecomments{(1) Gaussian component number. 
(2)--(3) Component peak coordinate.
(4) 2-D Gaussian axes FWHM. 
(5) Gaussian position angle.
(6) Radius defined as the geometric mean of the semi-major and semi-minor axes.
(7) Peak intensity from the Gaussian fit.
(8) Flux density.
(9) Brightness temperature measured from the peak intensity in the primary beam corrected continuum map.
(10) Median temperature from \methanol\ line fitting.
(11) Component gas mass. The value in brackets corresponds to $M_d^{\rm thin}$ estimated from the brightness temperature.
(12) Optical depth at peak intensity.
}
\end{deluxetable*}

\subsection{Physical properties}\label{sect:props}

\begin{figure*}[ht!]
\plotone{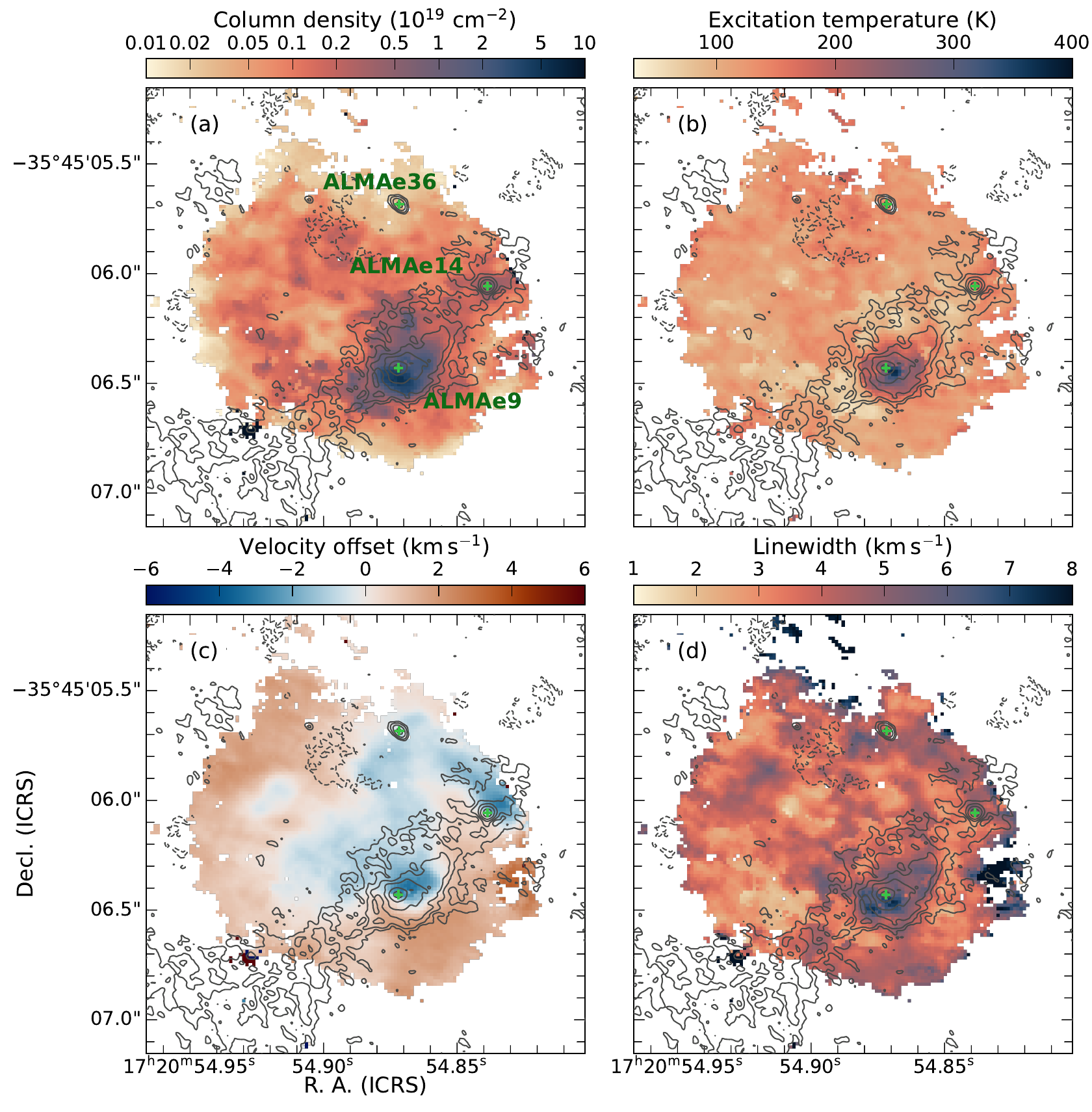}
\caption{Physical properties derived from LTE fitting of \methanol\ emission.
The continuum is shown in contour with the same levels as in Figure~\ref{fig:continuum}.
The zero velocity offset corresponds to a systemic velocity of --5.1\,\kms\ \citep{olguin_digging_2026}.
\label{fig:lte_fit}}
\end{figure*}

We use \methanol\ to calculate the physical properties of the gas, including the gas column density and temperature.
The fitting is performed with \textsc{spectuner}\footnote{\url{https://spectuner.readthedocs.io/en/latest/index.html\#}} \citep{qiu_spectuner_2025} under the local thermodynamic equilibrium assumption (LTE).
The Cologne Database for Molecular Spectroscopy \citep[CDMS;][]{endres_cologne_2016} is used by \textsc{spectuner} to find the lines covered in our observations.
A total of 19 \methanol\ transitions are available in the observations.
However, we exclude data below $4\sigma$ and pixels with less than 3 peaks in \textsc{spectuner}.
The resulting model cubes include 15 transitions with emission over $4\sigma$ (roughly 24\,K) close to the main sources.
To minimize the effects of outliers in the following calculations, we post-process the derived quantity maps.
First, we exclude pixels with emission below $5\sigma$ in the \methanol\ $E\,v_t=0\,J_{K_a,K_c}8_{{-0,8}}-7_{{-1,6}}$ line, which is the main line used in the analysis of the extended emission in the following sections.
Then, we use a Hampel test to search for outliers outside of a 3 median absolute deviation range in rolling boxes of size $5\times5$ pixels, and replace the outliers by the median value within each box.  
The results of the fit are shown in Figure~\ref{fig:lte_fit}.

The column density peaks to the south of ALMAe9, while the temperature peaks closer to the ALMAe9 continuum peak.
Near the continuum peak, optically thicker lines and strong continuum could result in non-Gaussian line profiles, thus the results near the continuum peak should be more uncertain (see Appendix~\ref{app:figs} for example line fits).
We estimate the median column density and temperature values within the regions defined by the Gaussian fit to the continuum emission (Table~\ref{tab:src_props}).
The difference between the brightness temperature measured at the peak continuum intensity of each source (Table~\ref{tab:src_props}) and the temperature from the \methanol\ fitting indicate that the continuum emission is not optically thick.

We use the median temperature, $T$, within the 2-D Gaussian regions fitted to the continuum emission to estimate the dust mass of the circumstellar gas as \citep[e.g.,][]{pouteau_alma-imf_2022}
\begin{equation}
     M_d = \tau_\nu \frac{\Omega_{\rm beam} F_{\nu} d^2}{R_{dg} \kappa_{\nu} I_\nu}\, = \tau_\nu M_d^{\rm thin} \frac{\Omega_{\rm beam} B_\nu(T)}{I_\nu}\,,
\end{equation}
where $\Omega_{\rm beam}$ is the beam area, $F_\nu$ and $I_\nu$ are the primary beam corrected flux density and peak intensity listed in Table~\ref{tab:src_props}, $d$ is the distance, $R_{dg}=0.01$ is the dust-to-gas ratio, $B_\nu (T)$ is the Planck blackbody function, and $M_d^{\rm thin}$ is the gas mass under the optically thin assumption. 
Following \citet{olguin_digging_2026}, we use the dust opacity law from \citet{yamamuro_impact_2025} which yields $\kappa_\nu = 0.24$\,cm$^2$\,g$^{-1}$ at 1.3\,mm.
The optical depth is defined as
\begin{equation}
    \tau_\nu = -\ln\left( 1 - \frac{I_\nu}{\Omega_{\rm beam} B_\nu(T)}\right)\,.
\end{equation}
The masses and optical depth values of each Gaussian component are listed in Table~\ref{tab:src_props}.
Using the ratio of the axes of the 2-D Gaussian components in Table~\ref{tab:src_props}, we estimate an inclination angle range between ${\sim}40-50\degr$, resulting in central source masses in the range of ${\sim}8-12$\,\msun\ from its uncorrected mass $M_c \sin^2 i=5$\,\msun\ \citep{olguin_digging_2026}.
However, given the extended nature of the continuum emission, these inclination angles may be less accurate than those found in a similar way in other sources \citep[e.g.,][]{wang_eccentric_2026}.
The total mass of the gas (1.4\,\msun) could help the central source grow further and, as discussed below, the circumstellar gas keeps being replenished from the core/circumbinary gas.
Note that the circumstellar gas mass is ${\gtrsim}10$\% that of the central sources, making the circumstellar gas presumably gravitationally unstable.
As expected, the median temperature and optical depths of the compact Gaussian component are higher than the more extended component. 
The difference between $M_d$ and $M_d^{\rm thin}$ ranges between $0.01-0.07$\,\msun\ ($4-23$\%).
Table~\ref{tab:src_props} also lists the value of $M_d^{\rm thin}$ calculated using the peak brightness temperature, which would correspond to an upper limit. 

\subsection{Gas kinematics}\label{sect:kinematics}

We use \methanol\ and SO emission to analyze the kinematics of the circumbinary/circumstellar gas.
For simplicity, we select the \methanol\ $E\,v_t=0\,J_{K_a,K_c}8_{{-0,8}}-7_{{-1,6}}$ (220.078561\,GHz; $E_u/k_B=97$\,K) and $A\,v_t=0\,J_{K_a,K_c}=18_{{3,15}}-17_{{4,14}}$ (233.795666\,GHz; $E_u/k_B=447$\,K) transitions, which hereafter will be referred to as cold and hot transitions based on their $E_u/k_B$. 
Similarly, we will refer to the SO $^3\Sigma\,v=1\,J_K=6_5-5_4$ (218.3238577\,GHz; $E_u/k_B=1634$\,K) transition as a hot gas and shock tracer.
The kinematics from cold gas tracers described below are observed in different tracers/transitions, we thus use these as additional support.
In particular, we use \metcyn\ $v=0\,J_K=12_3-11_3$ (220.7090165\,GHz; $E_u/k_B=133$\,K) as an additional cold transition to compare with previous works.

\subsubsection{Circumbinary gas motion}\label{sect:kinematics:circumbinary}

\begin{figure*}[ht!]
\epsscale{0.8}
\plotone{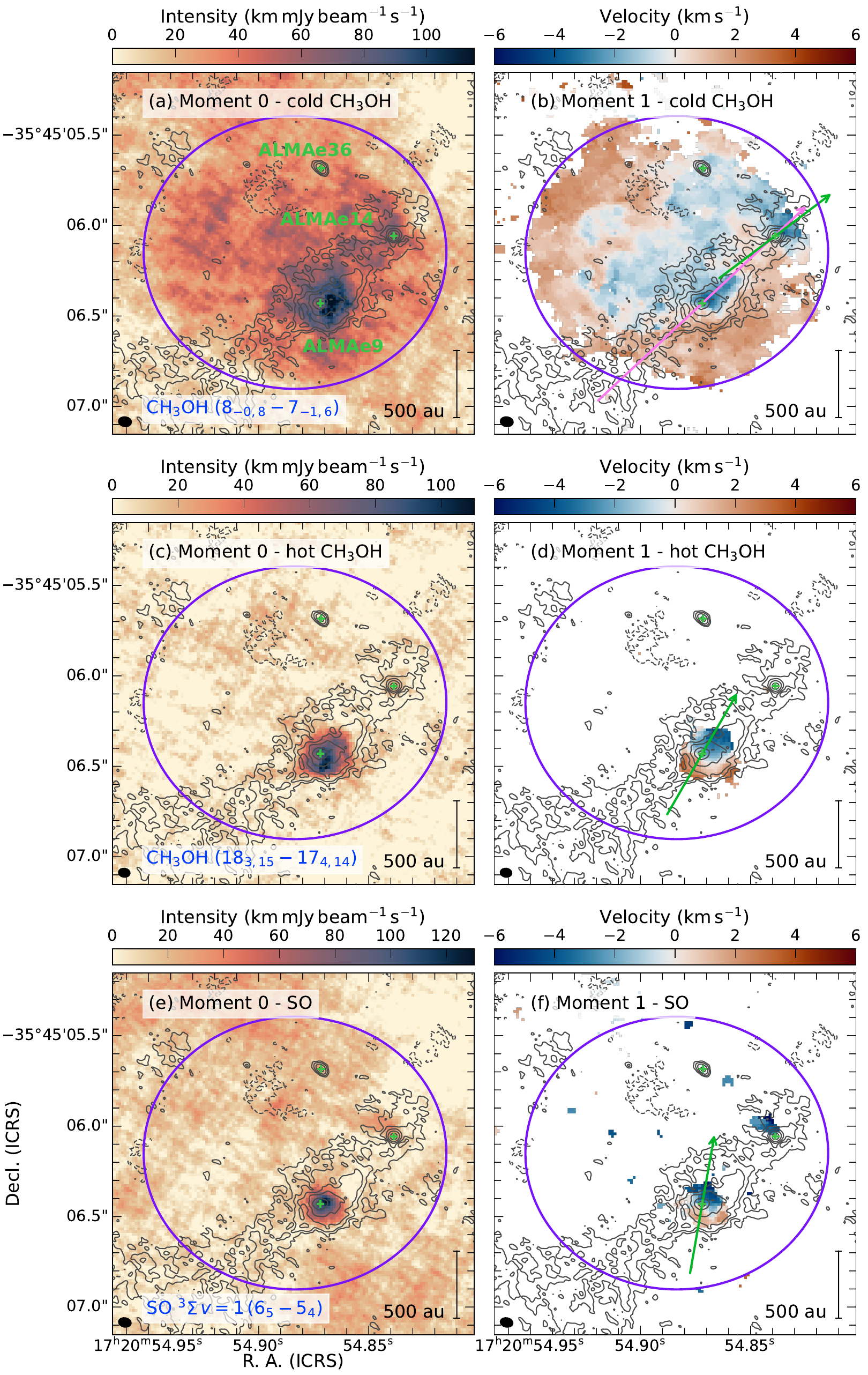}
\caption{Zeroth (left) and first (right) order moment maps from cold and hot gas tracers.
The molecular line transitions are labeled in the left column, and the velocities are shifted by the systemic velocity ($-5.1$\,\kms).
The moment maps are integrated within ${\pm}{\sim}10$\,\kms\ from the systemic velocity. 
The pink arrow indicates the direction of the PV map slit passing through both sources, while the green arrows indicate the direction of the velocity gradient close to the sources.
The purple circle indicate roughly the size of the molecular line emission over $5\sigma$ (radius ${\sim}0\farcs8{\sim}1000$\,au), which is the same threshold used to produce the first moment maps.
The 1.33\,mm continuum emission is shown in gray contours with the same levels as in Figure~\ref{fig:continuum}.
The synthesized beams of the moment maps are shown in the lower left corner.
\label{fig:moments}}
\end{figure*}

Figure~\ref{fig:moments} shows zeroth and first order moment maps of cold and hot transitions of \methanol\ as well as a vibrationally excited SO emission.
The gas emission traced by cold transitions of \methanol\  and other complex organic molecules (e.g., \metcyn) has a radius of roughly 0\farcs8 (1000\,au; see purple circle in Figure~\ref{fig:moments}, hereafter circumbinary gas) measured from emission over the $5\sigma$ level.
The sources are located off-center of this area.
Hence, the circumbinary gas could represent a gas reservoir with an anisotropic distribution and/or motion of the binary system with respect to the larger cloud/core.
This is also supported by the column density distribution in Figure~\ref{fig:lte_fit}(a).

\begin{figure*}[ht!]
\epsscale{0.8}
\plotone{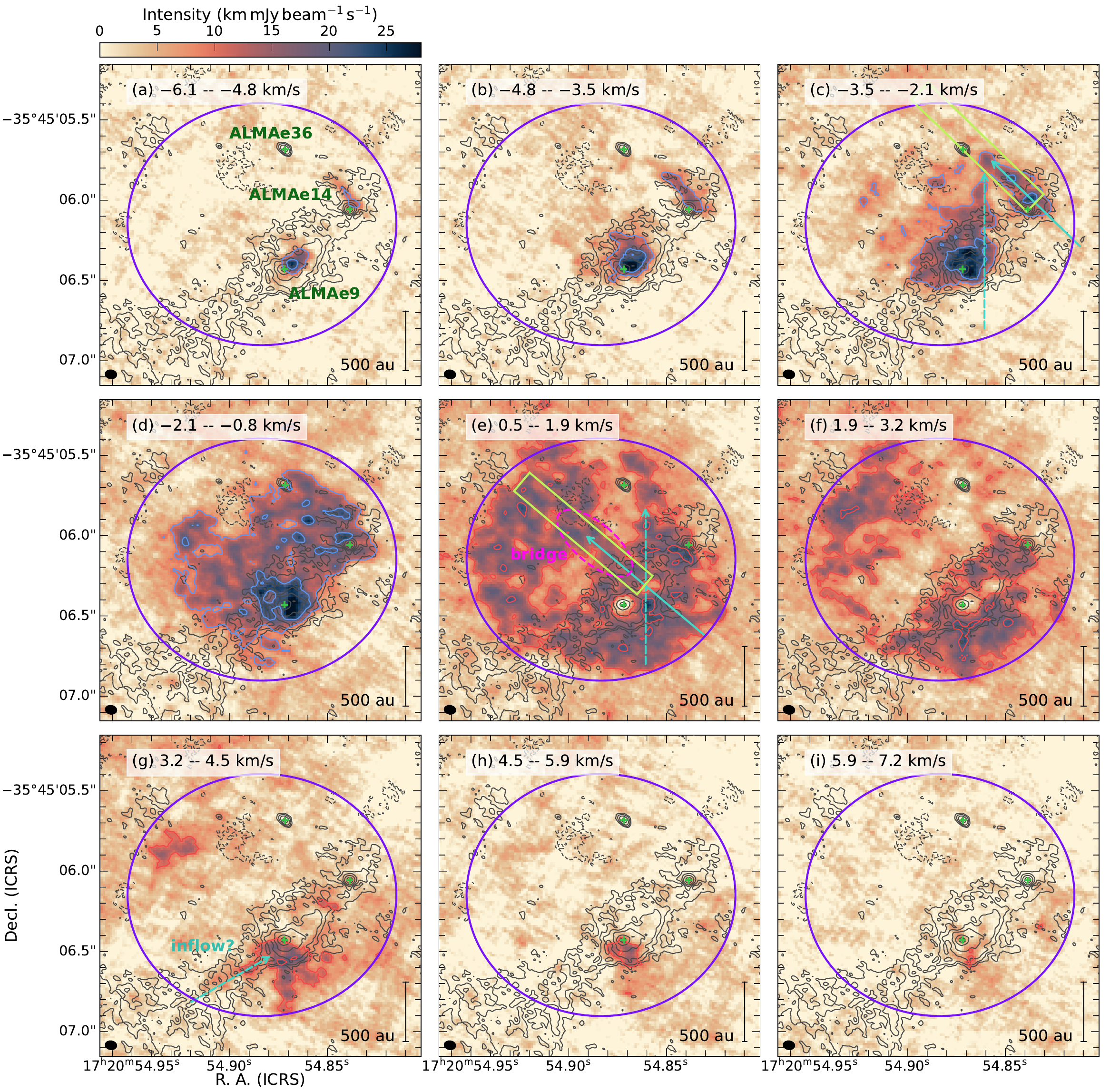}
\caption{Cold \methanol\ \metcold\ zeroth order moment maps at different velocity ranges (color map and blue/red contours).
The range of velocity offset for each map is shown in the upper left corner and are shifted by a systemic velocity of --5.1\,\kms.
The blue and red contour levels are 5, 10$\times\sigma$ with $\sigma=1$\,\mjybeam\,\kms, while the gray contours are the same continuum levels as in Figure~\ref{fig:continuum}. 
The purple circle indicate roughly the size of the molecular line emission over $5\sigma$, and the cyan arrows indicate the directions of the position-velocity maps in the potential inflows (see Section~\ref{sect:kinematics:circumbinary}).
The lime boxes correspond to the regions where average physical properties of the inflows are calculated in Section~\ref{sect:kinematics:inflows}.
The dashed pink ellipse in (e) indicate the bridge-like structure.
The cyan arrows in (c) and (e), and (g) indicate two additional inflow candidates.
The green crosses show the position of the continuum sources.
The synthesized beam ellipse is shown in the lower left corner.
\label{fig:chan_maps}}
\end{figure*}

The large scale gas shows complex kinematics.
To disentangle these motions, Figure~\ref{fig:chan_maps} show zeroth order moment maps calculated at different velocity ranges.
The maps show gas motion similar to that close to the sources from SE to NW, but also motion in the orthogonal direction toward the sources as the absolute value of the velocity increases (see also Figure~\ref{fig:lte_fit}(c)).
Thus the motions seem to be indicative of a combination of rotation and inflows.

\subsubsection{Circumstellar gas motions}\label{sect:kinematics:circumstellar}

The cold gas tracers close to the main continuum sources, ALMAe9 and ALMAe14, seem to be rotating in roughly the same direction (NW to SE) as indicated by the visually defined green arrows in Figures~\ref{fig:moments}(b) and (d).
This implies that their angular momentum vectors are oriented in a similar direction.
The extension of the hot transitions is more compact and located close to the sources.
The hot \methanol\ transition in Figures~\ref{fig:moments}(c) and (d) is detected mainly around ALMAe9, and only marginally detected in ALMAe14.
Around ALMAe9, vibrationally exited SO seem to be tracing the same gas motion as hot \methanol, albeit with a different velocity gradient P.A. (visually estimated as ${\sim}$330\degr\ for hot \methanol\ and ${\sim}350\degr$ for SO, see Figures~\ref{fig:moments}(d) and (f)).
This shift in angles may be explained by yet unresolved gas motions.
On the other hand, SO seems to be tracing a shock feature of a potential inflow around ALMAe14 (see Section~\ref{sect:kinematics:inflows}).

\begin{figure}[ht!]
\plotone{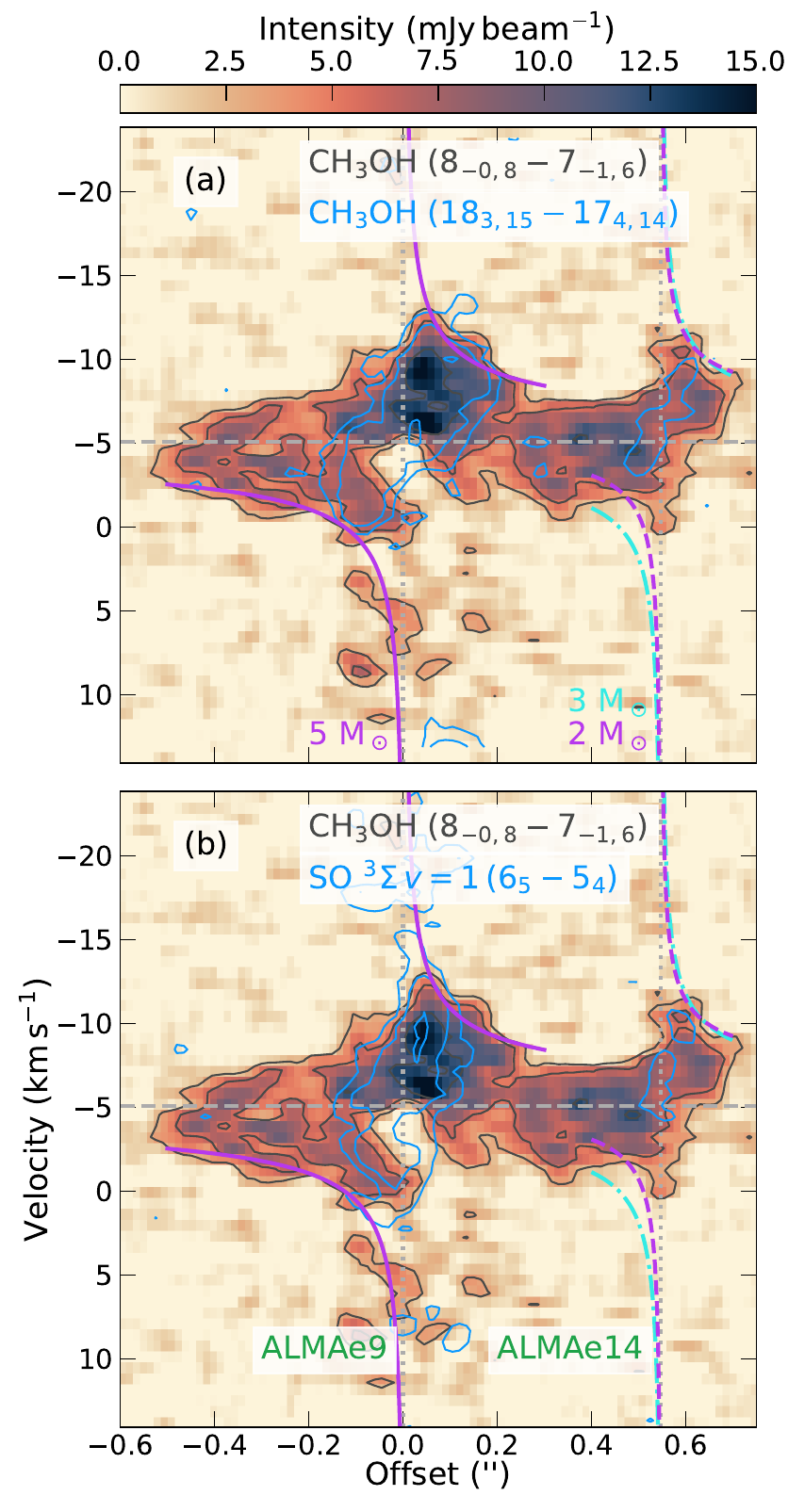}
\caption{Position velocity maps of cold (color scale and gray contours) and hot gas tracers (blue contours) for a slit passing through ALMAe9 and ALMAe14.
The contour levels are 3, 6, 12, ...$\times \sigma$, with $\sigma=1$\,\mjybeam.
The purple continuous line shows the Keplerian curve fitted to ALMAe9 from \metcyn\ emission in \citet{olguin_digging_2026}.
The cyan dot-dashed line shows the equivalent Keplerian curve from the upper limits to the mass in  \citet{olguin_digging_2026} centered at the same systemic velocity of ALMAe9 (--5.1\,\kms), while the purple dashed line shows an improved fit to the \methanol\ emission (see Section~\ref{sect:kinematics:circumstellar}).
The gray dotted vertical lines indicate the position of the sources, while the gray dashed horizontal one indicate the systemic velocity of ALMAe9. 
\label{fig:pv_common}}
\end{figure}

We calculate position velocity (PV) maps from a slit passing through both sources (P.A. = 313\degr;  cyan arrow in Figure~\ref{fig:moments}(b)).
The PV maps are shown in Figure~\ref{fig:pv_common}.
The PV map of the cold \methanol\ transition shows a typical distribution of rotation and infall with a red-shifted dip feature akin to inverse p-Cygni profiles close to the bright continuum source ALMAe9.
This map also shows emission connecting ALMAe9 and ALMAe14 with a slight velocity gradient from offsets larger than 0\farcs2.
In simulations, similar structures could allow gas to feed the primary through the secondary \citep[e.g.,][]{matsumoto_structure_2019}.

Around ALMAe14, the profile resembles rotation for blue-shifted velocities and larger offsets relative to the ALMAe14 location.
On the other hand, around ALMAe9 the hot gas tracers seem to be dominated by rotation motion, and their line profiles are less affected by the bright continuum source.
We show the rotation curves for ALMAe9 using the Keplerian curve from \citet[][]{olguin_digging_2026} in Figure~\ref{fig:pv_common}.
Note that they also fit a curve with a power law with the exponent as a free parameter, but the exponent (--0.69) is close to Keplerian (--0.5), thus the curves are relatively similar.
Although these curves are calculated to fit the PV map at slightly different P.A. (310\degr\ vs. 313\degr), they are in good agreement with the low level contours of the hot gas tracers.
For ALMAe14, \citet[][]{olguin_digging_2026} only estimated an upper limit for the mass based on the radius where the velocity peaks.
Using their values (mass of $M\sin^2i = 3$\,\msun, radius of 120\,au), we show the corresponding Keplerian curve in Figure~\ref{fig:pv_common}.
The lack of sensitivity of the hot gas tracers and the extended emission connecting the sources precludes a proper fit of the blue and red shifted contours around ALMAe14.
Therefore, we only vary the previous results to show a tentative curve that could fit the velocity profiles.
This results in an estimated systemic velocity of --6.2\,\kms\ and a mass of $M \sin^2 i\sim2$\,\msun\ within 120\,au.
The estimated inclination angle range is $20-40\degr$ from the ratio of the axes of the 2-D Gaussian components in Table~\ref{tab:src_props}, which yields a mass of $M{\sim}5-17$\,\msun.

\subsubsection{Potential inflows}\label{sect:kinematics:inflows}

Of the motions toward the sources, Figure~\ref{fig:chan_maps} shows an extended arm-like structure with bluer velocities in cold \methanol\ toward ALMAe14 (see the cyan arrow in Figure~\ref{fig:chan_maps}(c)).
The direction of the velocity increment (Figures~\ref{fig:chan_maps}(a)-(c)) indicates that this feature is likely an infalling structure, rather than an outflow which would require for the velocity to increase further from the source.
This feature is also similar to infalling streamers found in G336.01--0.82 \citep[][]{olguin_digging_2023,olguin_massive_2025} and IRAS 18134--1942 \citep{mai_misaligned_2025}.
Similar to \citet{olguin_massive_2025}, this arm-like feature is also seen in the first moment map of SO (Figure~\ref{fig:moments}(f)), which could imply the presence of accretion shocks \citep[e.g.][]{oya_infalling-rotating_2016}.
Alternatively, SO could be tracing an outflow shock while cold \methanol\ could be tracing infall in a surface layer of the outflow cavity walls \citep[e.g.,][]{tobin_observational_2024}.

A bridge-like structure is present in Figure~\ref{fig:chan_maps}(e) connecting the extended continuum emission of ALMAe9 with the outer regions of the circumbinary gas. 
The same extended continuum emission seem to be connected to filamentary-like \methanol\ emission to the north (dashed arrow in Figures~\ref{fig:chan_maps}(c) and (e)), which could trace an additional inflow splitting from or overlapping with the one feeding ALMAe14.
However, the PV map in Appendix~\ref{app:figs} does not show a clear velocity gradient.
On the other hand, extended red-shifted emission toward the SE coincides with the extended continuum emission in the same direction (Figure~\ref{fig:chan_maps}(g)).
A velocity gradient is observed in the PV map in the same direction (Figure~\ref{fig:pv_common} for negative offsets and red-shifted velocities), and may thus indicate infall from the core/circumbinary gas onto the primary source.
However, this velocity gradient can have a significant contribution from gas rotation.

\begin{figure*}[ht!]
\plotone{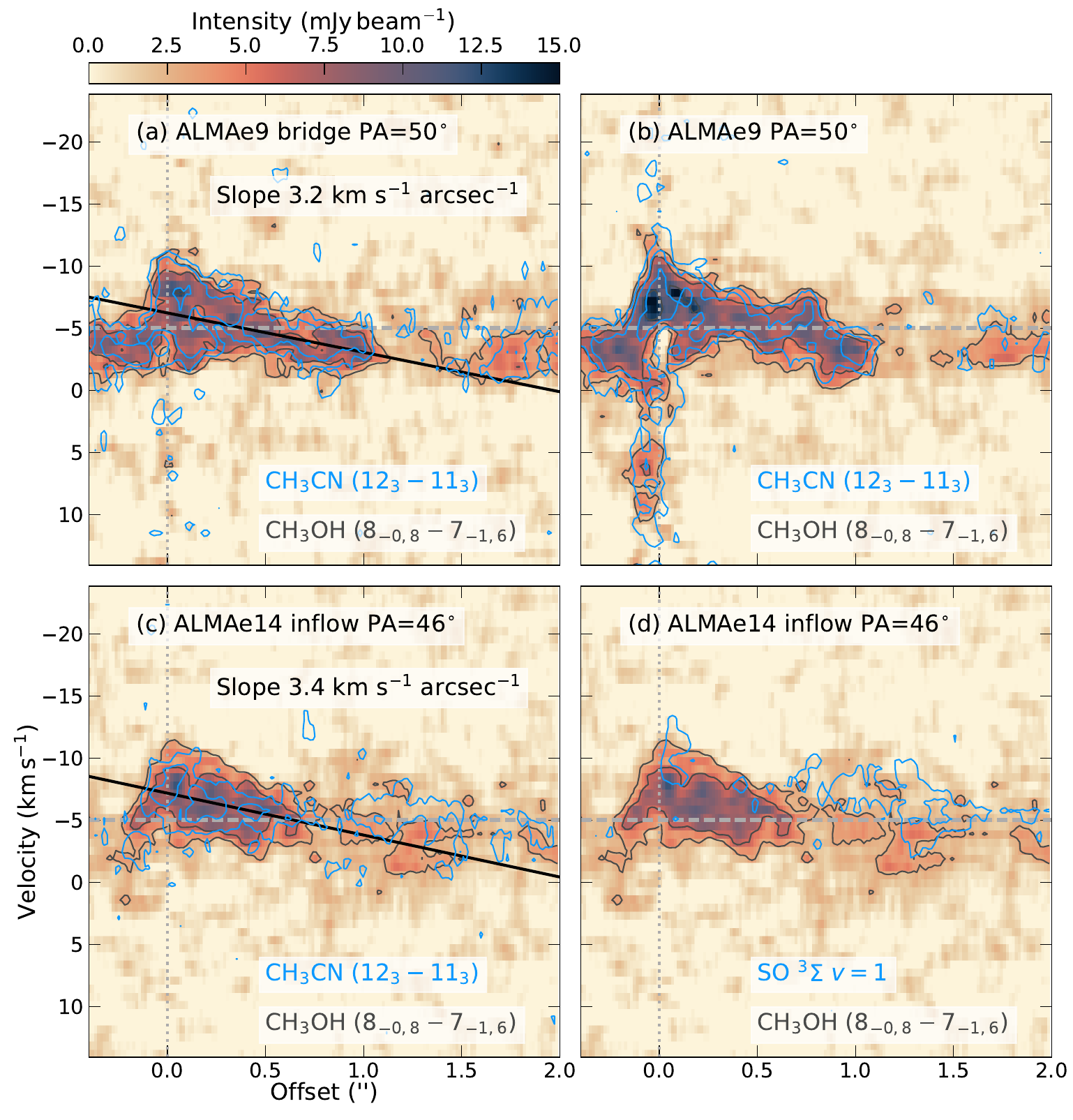}
\caption{PV maps in the direction of the potential inflows.
(a) PV map along the cyan arrow in Figure~\ref{fig:chan_maps}(e).
(b) PV in the same direction as (a) (same P.A.) but with its origin in ALMAe9.
(c)-(d) PV along the cyan arrow in Figure~\ref{fig:chan_maps}(c).
(a)(c)(d) The zero offsets are determined by the intersection between the pink arrow in Figure~\ref{fig:moments}(b) and the cyan arrows in Figures~\ref{fig:chan_maps}(e) and (c).
The black lines in panels (a) and (c) are fits to the intensity-average radial profile of the \methanol\ emission in color scale.
The contour levels are the same as those in Figure~\ref{fig:pv_common}.
The vertical and horizontal gray lines correspond to the zero offset and systemic velocity, respectively.
\label{fig:pv_flows}}
\end{figure*}

We explore the velocity gradients along the arrows in Figures~\ref{fig:chan_maps}(c) and (e), i.e., the arm-like structure near ALMAe14 and the bridge-like structure near ALMAe9.
Figure~\ref{fig:pv_flows} shows the PV maps along the aforementioned cyan arrows.
The cold molecular line emission from \methanol\ and \metcyn\ show similar linear velocity gradients for both sources.
We calculate an intensity weighted velocity profile from the cold \methanol\ emission and fit a linear function to find the velocity gradient.
For the velocity profile we use all the data over the $5\sigma$ level, and positive offsets up to 1\farcs0 for ALMAe14, while for ALMAe9 we select offsets between 0\farcs4 and 1\farcs1.
These are mostly enclosed within the extension of the \methanol\ emission (purple area in Figure~\ref{fig:chan_maps}). 
We obtain a velocity gradient, $\nabla v$, of 3.2 and 3.4\,\kms\,arcsec$^{-1}$ for the inflows close to ALMAe9 and ALMAe14, respectively.
To estimate the infall rate, we first calculate an average \methanol\ column density, $N({\rm CH_3OH})$, from Figure~\ref{fig:lte_fit}(a) in the box regions shown in Figure~\ref{fig:chan_maps}.
The length of these boxes represent the extension of the inflows (1\farcs0), while width roughly matches the width of the inflows from the molecular line emission (0\farcs15). 
We assume a \methanol\ abundance range of $N({\rm CH_3OH})=10^{-5}-10^{-6}N({\rm H_2})$, which is representative of gas over 100\,K in hot cores \citep[][]{li_alma_2025}, to estimate the gas mass of the inflows, $M_{\rm in}$, as \citep[e.g.,][]{valdivia-mena_prodige_2022}
\begin{equation}
    M_{\rm in} = \mu m_H d^2 \Omega_{\rm pix} \sum N({\rm H_2})\,,
\end{equation}
where $\Omega_{\rm pix}$ is the pixel angular area, and $\mu=2.8$ is the mean molecular weight \citep[][]{kauffmann_mambo_2008}. 
This results in inflow gas masses of $3.2\times(10^{-2}-10^{-1})$\,\msun\ for ALMAe9 and $7.1\times(10^{-3}-10^{-2})$\,\msun\ for ALMAe14.
We estimate an infall rate assuming a cylindrical inflowing filament \citep[][]{kirk_filamentary_2013}
\begin{equation}
    \dot{M} = \frac{\nabla v M_{\rm in}}{\tan \alpha}
\end{equation}
where $\nabla v$ is the velocity gradient and $\alpha$ is the angle between the filament axis and the plane of the sky.
Assuming $\alpha=45\degr$, the infall rate onto ALMAe9 is $1.6\times(10^{-4}-10^{-5})$\,\msun\,yr$^{-1}$, while for ALMAe14 is $3.8\times(10^{-5}-10^{-6})$\,\msun\,yr$^{-1}$.
From the standard deviation of the mean of the column density values and propagation of error, we estimate an uncertainty of ${\sim}5\%$ in these infall rate estimations.
This estimation does not considers the flux calibration error, estimated to be of about 10\%.
These results indicate that the circumstellar structures are fed at similar rates.
In the simulations of \citet[][]{matsumoto_structure_2019}, similar infall rates toward the primary and secondary are obtained in their simulations with lower angular momentum.

Figure~\ref{fig:pv_flows}(b) shows a PV map passing through the source ALMAe9 with the same P.A. as the off-center map shown in Figure~\ref{fig:pv_flows}(a).
However, unlike the off-center map, the PV map shows a relatively flat profile between 0\farcs4 and ${\sim}0\farcs8$ where a kink is observed before a sudden increase in velocity.
On the other hand, the SO emission toward ALMAe14 is offset in velocity.
Given that SO is detected closer to the source than the PV slit, this may be a marginal detection of gas at higher velocity near the source. 
Otherwise, the SO emission may point  to a different velocity component or contamination from other lines in the same frequency range.

\section{Conclusions}\label{sect:conclusions}

We observed the proto-binary system NGC 6334I(N) ALMAe9-ALMAe14 with ALMA at 1.33\,mm with a resolution of ${\sim}60$\,mas (${\sim}90$\,au).
The primary and secondary sources are separated by 740\,au and surrounded by circumbinary gas of size of roughly 1000\,au revealed by molecular line emission, with the system off-center of the circumbinary gas.
The observations resolve the circumstellar emission in continuum and hot gas tracers.

Using the \methanol\ emission, we calculated temperature and column density maps.
By combining the temperature and continuum emission, we estimate masses of 1.6 and 0.6\,\msun\ for the circumstellar gas of the primary and secondary, respectively.

Cold transitions of \methanol\ also reveal rotation and infalling motions feeding gas from circumbinary to circumstellar scales.
The tentative rotation profile of the secondary implies an enclosed/stellar mass of 2\,\msun, assuming Keplerian rotation and not corrected for source inclination.
The hot gas tracers show similar rotation of the circumstellar gas around the primary as the cold gas tracer.
The rotation axes of the primary and secondary are roughly pointing in the same direction, implying that they likely formed relatively close due to disk fragmentation.  

We identify two potential inflows and estimated infall rates of $10^{-4}-10^{-6}$\,\msun\,yr$^{-1}$, with the primary and secondary circumstellar gas fed at similar infall rates.
These inflows may allow the stellar masses to continue growing, eventually forming a binary system with at least one high-mass star.

\begin{acknowledgments}
F.O. and Y.O. acknowledge the support of the NAOJ ALMA Joint Scientific Research Program grant No. 2024-27B.
Y.O. acknowledges the support by Grant-in-Aids from Ministry of Education, Culture, Sports, Science, and Technologies of Japan (KAKENHI; 21K13954, 25K07367).
P.S. was supported by the NAOJ ALMA Scientific Research Grant Code 2026-31B and Grant-in-Aid for Scientific Research (KAKENHI Number JP26H02066, JP23H01221, and JP26K00748) of JSPS.
K.T. is supported by JSPS KAKENHI grant Nos. 21H01142, 24K17096, and 24H00252.
H.V.C. acknowledges financial support from National Science and Technology Council (NSTC) of Taiwan grants NSTC 114-2112-M-007-043.
This paper makes use of the following ALMA data: ADS/JAO.ALMA\#2016.1.01036.S and \#2022.1.00700.S. ALMA is a partnership of ESO (representing its member states), NSF (USA) and NINS (Japan), together with NRC (Canada), NSTC and ASIAA (Taiwan), and KASI (Republic of Korea), in cooperation with the Republic of Chile. The Joint ALMA Observatory is operated by ESO, AUI/NRAO and NAOJ.
The Scientific color maps vik and lipari \citep{crameri_scientific_2023} are used in this work to prevent visual distortion of the data and exclusion of readers with color vision deficiencies \citep{crameri_misuse_2020}.
\end{acknowledgments}

\facilities{ALMA}

\software{CARTA \citep{wang_cartacube_2026,angus_comrie_carta_2026},
          CASA \citep{casa_team_casa_2022},
          astropy \citep{astropy_collaboration_astropy_2022},
          astropy-regions \citep{larry_bradley_astropyregions_2024},
          spectuner \citep{qiu_spectuner_2025}
          }

\appendix

\section{Supplementary figures}\label{app:figs}

\begin{figure}[ht!]
\epsscale{0.6}
\plotone{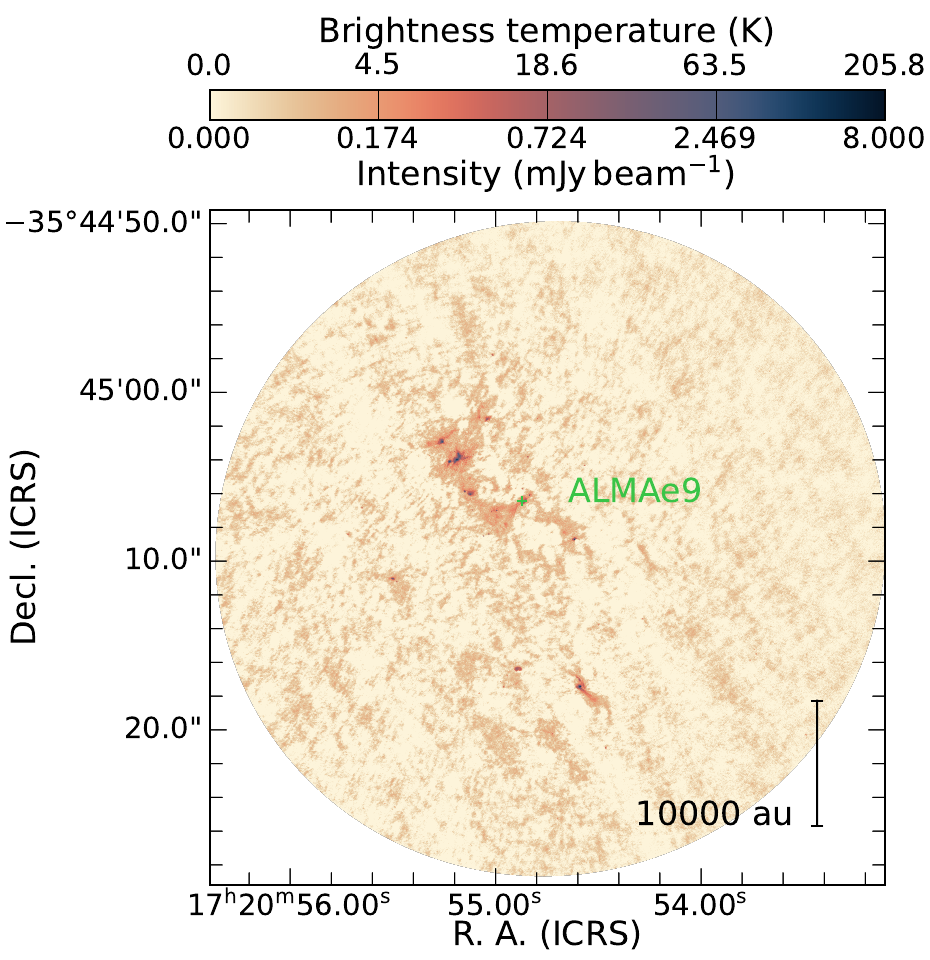}
\caption{ALMA 1.3 mm continuum emission of the NGC 6334I(N) field. 
The green cross marks the position of the ALMAe9 source.
\label{fig:field}}
\end{figure}

\begin{figure}[ht!]
\epsscale{0.8}
\plotone{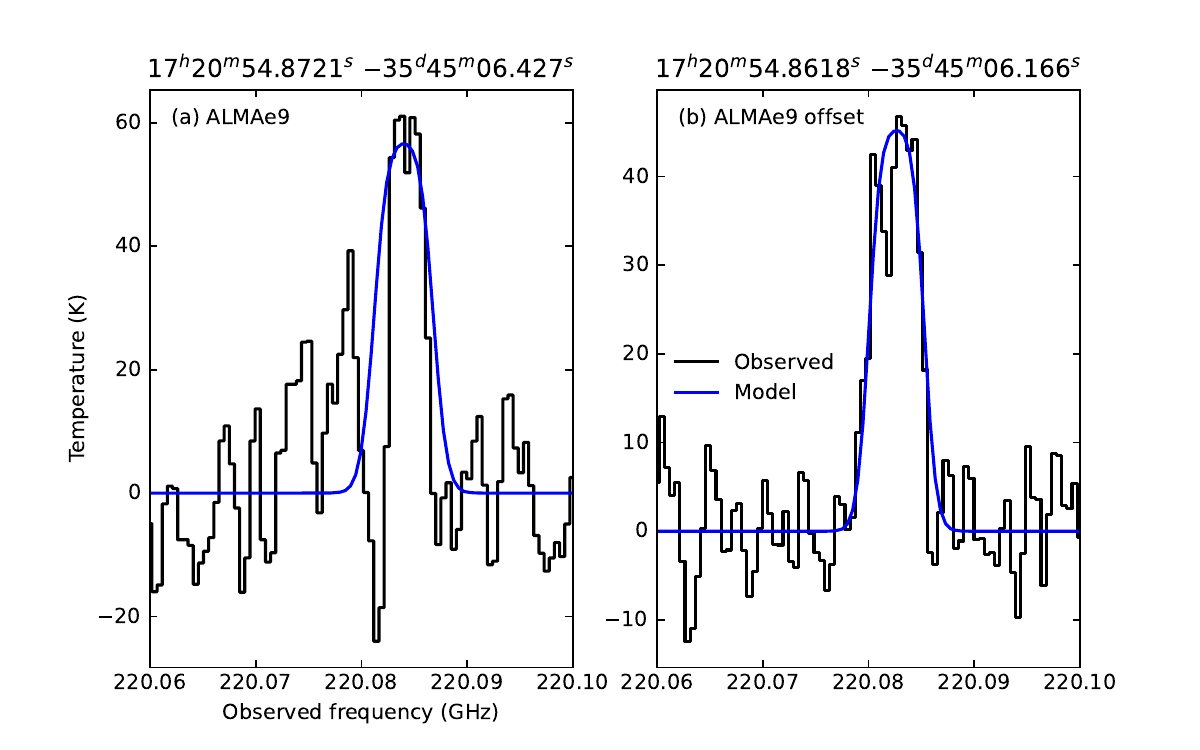}
\caption{Example cold \methanol\ \metcold\ spectra and LTE fitting at the position of ALMAe9 and an offset position.
\label{fig:ex_spec}}
\end{figure}

\begin{figure}[ht!]
\epsscale{0.8}
\plotone{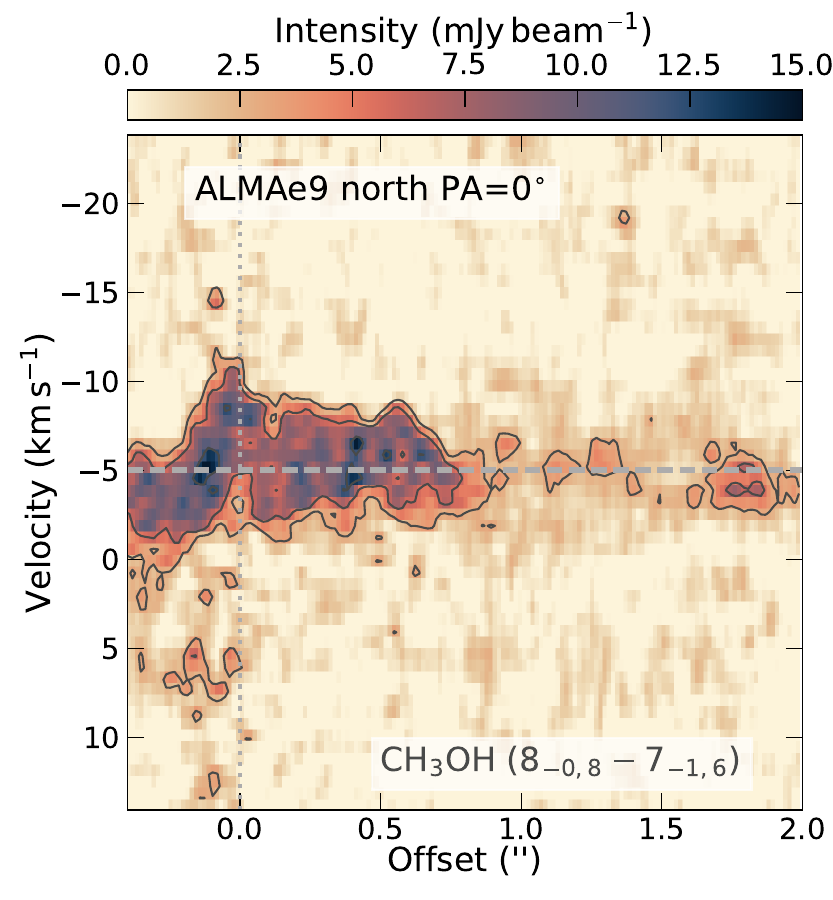}
\caption{PV map \methanol\ \metcold\ along the apparent inflow in the direction of the dashed arrows in Figures~\ref{fig:chan_maps}(c) and (e).
\label{fig:north_flow}}
\end{figure}

Figure~\ref{fig:field} shows the NGC 6334I(N) field.
Figure~\ref{fig:ex_spec} shows example spectra of the observed and model cold \methanol\ emission.
Figure~\ref{fig:north_flow} shows the PV map along the dashed arrows in Figures~\ref{fig:chan_maps}(c) and (e).

\bibliography{references_fixed}{}
\bibliographystyle{aasjournalv7}

\end{document}